# Calibrated selective prediction using deep ensembles for thyroid nodule ultrasound classification under dataset shift

Md. Sadibul Hasan Sadib [a], Md. Mohayminul Mukit [a], Rahmatul Kabir Rasel Sarker [a,*], Tahmid Alam Tamim [a], Md. Nazmul Hasan [a], Md. Monir Hossain Shimul [b]

[a] *Department of Computer Science and Engineering, Daffodil International University, Dhaka, Bangladesh*
[b] *Department of Public Health, Daffodil International University, Dhaka, Bangladesh*

ARTICLE INFO

ABSTRACT

*Background:* Deep learning can accurately classify thyroid nodules on ultrasound, but discrimination alone does not establish clinical reliability. We developed a calibrated deep-ensemble framework incorporating uncertainty-aware selective prediction and conservative image-based triage.

*Methods:* Backbone selection used nested five-fold cross-validation with three-fold inner architecture selection. ConvNeXt-Tiny with a squeeze-and-excitation block was selected consistently and used to construct a deterministic five-member ensemble. Within each outer-development partition, images were separated into training, early-stopping, calibration, and threshold-selection subsets, while the outer-test data remained isolated. Mutual information (MI), predictive variance, and maximum-softmax-probability (MSP)-derived uncertainty were compared using risk–coverage and triage analyses. Predictive values were assessed across assumed malignancy prevalences of 5%–70%. External evaluation applied the frozen TN5000-derived ensemble and operating parameters to TN3K without recalibration.

*Results:* On TN5000, pooled out-of-fold AUC-ROC was 0.9377, average precision 0.9708, expected calibration error 0.0138, and Brier score 0.0827. Under the primary MI-guided policy, 8.86% of images received a No-FNA suggestion, 60.00% an FNA recommendation, and 31.14% radiologist review. Among 3574 malignant images, 14 (0.39%) received a No-FNA suggestion and 708 (19.81%) were assigned for review. Predictive variance and MSP-derived uncertainty showed better error-ranking performance than MI on AURC and AUGRC, although no measure consistently optimized all triage outcomes. Across malignancy prevalences of 5%–70%, NPV ranged from 99.95% to 97.93%, while PPV increased from 29.28% to 94.75%. On TN3K, the frozen policy achieved an AUC-ROC of 0.7870, average precision of 0.7254, expected calibration error of 0.1899, and an 83.7% radiologist-review rate.

*Conclusion:* The ensemble demonstrated strong internal discrimination and calibration with conservative selective prediction. However, performance and calibration deteriorated substantially on TN3K, indicating limited transportability under dataset shift. Multicentre validation, site-specific recalibration, threshold validation, and prospective workflow evaluation are needed before clinical implementation.

## 1. Introduction

Thyroid nodules are frequently detected during routine clinical evaluation, particularly with widespread use of high-resolution ultrasonography. Although most nodules are benign, reliable risk stratification is essential to identify lesions that may warrant fine-needle aspiration (FNA) biopsy or further diagnostic assessment [1]. Current management is guided primarily by ultrasound-based frameworks, including the American Thyroid Association (ATA) guidelines and the American College of Radiology Thyroid Imaging Reporting and Data System (ACR TI-RADS) [1,2]. These systems integrate sonographic characteristics, nodule size, and clinical context to support surveillance and biopsy decisions. However, interpretation remains dependent on reader expertise, and substantial overlap between benign and malignant imaging appearances can complicate risk assessment.

Deep learning has shown considerable promise for thyroid nodule classification from ultrasound images. Early convolutional neural network studies demonstrated that image-based models could differentiate benign from malignant nodules with performance comparable to experienced clinicians in selected cohorts [3,4]. Subsequent research has explored transfer learning, multiview imaging, multimodal ultrasound fusion, multicentre decision-support systems, and comparative evaluation of modern neural architectures [5–12]. Together, these studies indicate that deep learning can extract clinically relevant imaging representations and potentially support radiologists during thyroid nodule assessment.

* Corresponding author. Department of Computer Science and Engineering, Daffodil International University, Savar, Dhaka, Bangladesh.
*E-mail address:* raselsarker.cse@diu.edu.bd (R.K.R. Sarker).

Nevertheless, high discrimination alone is insufficient for safe clinical decision support. A model may produce a confident prediction for an ambiguous image, for a case insufficiently represented in the development data, or for an image acquired under a different protocol or device. In such settings, forcing an automated classification may be inappropriate. A clinically cautious system should therefore identify cases for which an automated recommendation is unreliable and defer them to radiologist review.

Probability calibration and uncertainty estimation are central to this objective. Calibration assesses whether predicted probabilities correspond to observed outcome frequencies and is particularly important when probabilities are used to define conservative rule-out or rule-in thresholds [13]. Uncertainty estimation provides an additional signal for identifying predictions that should not be treated as stand-alone recommendations. In medical imaging, uncertainty-aware approaches commonly distinguish uncertainty related to image ambiguity from uncertainty arising from limited model knowledge or incomplete training-data coverage [14,15]. Deep ensembles and Monte Carlo Dropout are widely used practical approaches for estimating predictive variability across multiple model outputs [16,17]. However, both calibration and uncertainty behavior may deteriorate under dataset shift, even when discrimination remains acceptable [18].

Selective prediction, also known as classification with a reject option, offers a practical framework for incorporating these considerations into decision support by trading coverage for lower retained-case risk. Rather than requiring a model to issue an automated classification for every image, selective systems retain lower-risk predictions and defer uncertain cases to a human expert or an alternative diagnostic pathway [19–21]. This paradigm is especially relevant to thyroid ultrasound, where automated predictions should support, rather than replace, radiologist assessment.

The interpretation of triage performance depends on the prevalence and case mix of the evaluation cohort. TN5000 is a malignancy-enriched image dataset, whereas TN3K has a substantially different class distribution and annotation structure [22,23]. Consequently, predictive values and pathway-assignment rates cannot be assumed to transfer directly to unselected clinical populations or different sites. We therefore recalculated NPV and PPV across assumed malignancy prevalences while holding the observed class-specific performance fixed and interpreted these estimates as sensitivity analyses rather than evidence of clinical transportability. Accordingly, the evaluation considered probability calibration, uncertainty-based deferral, prevalence sensitivity, and the transportability of a frozen operating policy.

Few thyroid ultrasound studies have jointly examined probability calibration, ensemble-disagreement-based referral, conservative selective triage, and whether a fixed operating policy remains valid on an external dataset without recalibration or threshold adjustment.

In this study, we present a calibrated deterministic deep-ensemble framework for annotation-assisted thyroid nodule region-of-interest classification and selective image-based triage. The framework combines a ConvNeXt-Tiny backbone with squeeze-and-excitation attention, five independently trained ensemble members, member-wise vector-scaling calibration, and mutual information as an ensemble-disagreement score for deferral. Each image is assigned to one of three retrospective pathways: an image-based No-FNA suggestion, an image-based FNA recommendation, or radiologist review. Automated suggestions are issued only for low-disagreement cases with calibrated probabilities in conservative rule-out or rule-in regions, whereas cases with elevated ensemble disagreement or intermediate probabilities are deferred to radiologist review.

The contributions of this study are as follows:

1. We develop a calibrated deterministic deep-ensemble framework for annotation-assisted thyroid nodule ROI malignancy classification using ConvNeXt-Tiny with squeeze-and-excitation attention.
2. We evaluate mutual information as an ensemble-disagreement score for selective prediction and use it to identify images unsuitable for automated image-based triage suggestions.
3. We propose a conservative three-tier retrospective triage policy that combines uncertainty-based deferral with probability-based No-FNA suggestion and FNA-recommendation pathways.
4. We evaluate a frozen TN5000-derived calibration and triage policy on the independent TN3K cohort to assess discrimination, calibration, uncertainty behavior, and threshold transportability under dataset shift.

The objective of this study is not to replace clinical assessment or estimate real-world biopsy avoidance. Rather, this study evaluates whether calibrated ensemble disagreement can support a conservative selective-prediction framework in which automated image-based suggestions are issued only for low-disagreement cases, while potentially unreliable predictions are withheld and assigned to expert review.

## 2. Related work

Prior thyroid ultrasound studies have shown that deep learning can support benign-versus-malignant nodule classification and, in some settings, assist clinical management. However, clinical translation requires more than discrimination alone. Reliable decision support also requires calibrated probabilities, uncertainty-aware referral, and evidence that operating thresholds remain meaningful under dataset shift. Table 1 summarizes representative studies and positions the present framework.

**Table 1**
Representative studies in thyroid ultrasound deep learning and the positioning of the present study.

| Study | Primary focus | Reliability or clinical relevance | Relevance to the present study |
|---|---|---|---|
| Guan et al. [3] | Ultrasound-based benign-versus-malignant classification | Large-scale CNN classification | Established early evidence for deep-learning-based thyroid nodule classification |
| Peng et al. [4] | Multicentre diagnosis and management support | Clinical decision-support evaluation | Demonstrated potential for AI-assisted thyroid nodule management |
| Weng et al. [10] | Independent-dataset validation | Evaluation across different ultrasound equipment | Highlights the importance of external generalization |
| Ni et al. [11] | AI-assisted reduction of unnecessary FNA | Management-oriented retrospective evaluation | Illustrates the potential clinical value of AI-supported triage |
| Rashed et al. [12] | Comparative ultrasound classification | Deterministic deep-learning benchmarking | Provides recent evidence for conventional image-based classification |
| Saini et al. [24] | Multimodal ultrasound classification | Uncertainty-aware multimodal prediction | Demonstrates the value of reliability-aware prediction |
| Xiang et al. [25] | Segmentation and TI-RADS classification | Uncertainty-weighted multitask learning | Uses uncertainty during training rather than for case-level referral |

### 2.1. Deep learning for thyroid ultrasound classification and management

Deep-learning models can learn discriminative imaging patterns directly from thyroid ultrasound data and have demonstrated promising performance for benign-versus-malignant classification. Early work by Guan et al. [3] established the feasibility of convolutional neural networks for this task. Subsequent studies extended this direction through transfer learning, multiview imaging, multimodal ultrasound fusion, and modern neural architectures [5–8], [12,26].

Research has also moved toward clinical decision support. Peng et al. [4] developed a multicentre AI system for thyroid nodule diagnosis and management, and Ni et al. [11] evaluated an AI-assisted strategy to reduce unnecessary FNA procedures. These studies show the clinical potential of image-based AI. However, biopsy and surveillance decisions also depend on ultrasound features, nodule size, clinical history, and radiologist judgement. Image-based triage systems should therefore separate automated suggestions from cases requiring expert review, especially when biopsy-related recommendations are inferred from ROI-level predictions rather than full clinical assessment.

### 2.2. Calibration, uncertainty, and selective referral

High discrimination does not guarantee reliable probability estimates. A model may achieve a strong AUC while producing probabilities that do not reflect observed malignancy frequencies. Calibration is therefore essential when probability thresholds are used for rule-out or rule-in decisions [13]. In medical imaging, uncertainty estimation provides an additional safeguard by identifying cases for which an automated prediction may be unreliable [14,15].

Deep ensembles estimate prediction reliability through disagreement among independently trained models [16], whereas Monte Carlo Dropout estimates uncertainty using repeated stochastic inference [17]. However, uncertainty and calibration may deteriorate under dataset shift, even when discrimination remains useful [18]. Selective prediction addresses this issue by allowing a model to defer uncertain cases for expert review instead of forcing an automated decision for every image [19,20].

Thyroid-specific uncertainty-aware studies remain limited. Saini et al. [24] incorporated uncertainty into multimodal ultrasound prediction, while Xiang et al. [25] used uncertainty weighting for multitask learning. These approaches are valuable but differ from an inference-time selective-referral policy, where uncertainty is used after model training to determine whether a specific case should receive a non-review image-based suggestion or be assigned to radiologist review.

### 2.3. Cross-domain evidence on robust and interpretable artificial intelligence

Recent nonmedical engineering studies provide cross-domain evidence on ensemble design, robustness, and interpretability. Alpsalaz et al. [27] combined heterogeneous regressors within a stacking framework for photovoltaic power estimation and fault detection, whereas Özüpak et al. [28] evaluated a multilevel stacking framework for photovoltaic monitoring. Özüpak [29] integrated gated recurrent unit representations with XGBoost for lithium-ion battery state-of-health estimation using time-aware evaluation and post hoc explanations. Özüpak and Du [30] examined CNN-GRU remaining-useful-life prediction for turbofan engines under a reliability-oriented protocol, and Özüpak et al. [31] evaluated recurrent aircraft-component fault diagnosis under dynamically varying operating conditions using SHAP and LIME.

These studies illustrate transferable methodological principles: ensemble or hybrid improvements should be compared with appropriate baselines, robustness should be examined under operational variation, and model explanations should be evaluated separately from predictive accuracy. Their direct clinical relevance is limited because they do not evaluate thyroid imaging, clinical cohorts, probability calibration, prevalence-dependent predictive values, selective classification, or radiologist referral. The aircraft fault-diagnosis study used maximum-softmax confidence thresholds to flag low-confidence predictions, while identifying formal uncertainty-quantification methods as future work [31].

Explainability, calibration, and uncertainty therefore address distinct reliability questions. SHAP and LIME can identify input features associated with a prediction, but they do not establish whether predicted probabilities correspond to observed outcome frequencies or whether an automated output should be withheld. The present study addresses the latter questions through member-wise calibration, ensemble-disagreement analysis, selective prediction, fold-specific threshold evaluation, and assessment under external dataset shift.

Based on these gaps, we evaluated a calibrated deterministic five-member ensemble for annotation-assisted thyroid nodule ROI classification. MI-based ensemble disagreement deferred uncertain cases, while calibrated probability thresholds assigned retrospective image-based No-FNA suggestions or FNA recommendations only among low-disagreement cases. The frozen TN5000-derived ensemble and operating policy were subsequently evaluated on TN3K to assess transportability under dataset shift.

## 3. Methodology

### 3.1. Datasets and study design

This retrospective study used two publicly available thyroid ultrasound datasets: TN5000 for model development and internal evaluation, and TN3K for external domain-shift evaluation [22,23]. TN5000 supported nested architecture selection, five-fold internal evaluation. TN3K remained untouched during backbone selection, model fitting, calibration, uncertainty-score selection, or operating-threshold determination.

TN5000 comprised 5000 B-mode thyroid ultrasound images with binary benign or malignant labels and nodule-localization bounding boxes [22]. The analyzed cohort contained 1426 benign images (28.5%) and 3574 malignant images (71.5%); all internal analyses used these verified labels and counts.

The TN5000 data descriptor reports that one representative image was retained from the same patient perspective during dataset curation [22]. However, the released files contain neither patient identifiers nor an image-to-patient mapping; patient-level uniqueness and independence therefore could not be verified. Internal resampling was consequently performed at the image level. Before fold construction, SHA-256 hashing confirmed 5000 unique file-content hashes among 5000 images, with no exact duplicate groups. Difference-hash screening identified no candidate image pair within a Hamming distance of 2. Automated disjointness assertions were applied to the outer and inner partitions and were configured to stop execution if any sample identifier, SHA-256 hash, or leakage group crossed a split boundary. These controls exclude exact and detected near-duplicate image reuse but cannot exclude distinct images from the same patient when patient linkage is unavailable.

TN3K was evaluated using the official test split released with the dataset descriptor, comprising 614 images, including 378 benign images (61.6%) and 236 malignant images (38.4%) [23]. Although TN3K had a lower malignant proportion than TN5000, its class distribution should not be interpreted as representative of a population-screening cohort.

The datasets also differed in lesion-localization annotations and ROI-construction procedures. TN5000 supplied bounding-box annotations, from which lesion-centred ROI crops were extracted, whereas TN3K supplied pixel-wise segmentation masks used to derive lesion-centred ROI crops. Because both datasets provided lesion-localization annotations, this study evaluates annotation-assisted ROI classification rather than end-to-end nodule detection. Dataset-specific ROI-construction and input-standardization procedures are detailed in Supplementary

Methods S1. Differences in data source, class distribution, annotation granularity, and ROI construction were treated as components of the external dataset shift when interpreting TN3K performance. Table 2 summarizes the datasets, and representative extraction and ROI examples are provided in Supplementary Figs. S1–S2.

### 3.2. Overall framework

The proposed framework is summarized in the workflow schematic shown in Fig. 1. Across the five outer folds, each 1000-image partition served once as the untouched outer-test set, while the remaining 4000 images formed the corresponding outer-development set. Nested architecture selection and all fold-specific model-development procedures were restricted to the outer-development data. TN3K is presented as a separate external-evaluation branch in which the frozen TN5000-derived ensemble, calibration parameters, and operating thresholds were applied without recalibration or threshold adjustment. Grad-CAM is shown as a separate dashed qualitative-interpretation branch generated from one trained ensemble member because it did not contribute to ensemble prediction, uncertainty estimation, or triage assignment.

For each of five stratified outer folds, 1000 images (20%) were reserved as an untouched outer test partition before any architecture comparison. The remaining 4000 images formed the outer-development set. Within each outer-development set, ConvNeXt-Tiny, EfficientNetV2-S, DenseNet121, ResNet18, and MobileNetV4 were compared using three-fold inner cross-validation. Each inner fold used a rotating architecture-validation partition, while model fitting and early stopping were confined to the remaining inner-development data. Mean inner-fold AUC-ROC was the primary ranking criterion, with mean inner-fold average precision as the secondary criterion.

ConvNeXt-Tiny ranked first in all five outer folds. Following fold-specific architecture selection, the selected architecture was trained and calibrated using only the corresponding outer-development data and evaluated once on the untouched 1000-image outer test partition. Five ConvNeXt-Tiny models with different random initializations were trained within each fold. Member-wise vector scaling was fitted using the fold-specific calibration subset only. Calibrated member probabilities were aggregated to obtain the ensemble-mean malignancy probability and mutual information (MI).

The triage policy was applied in two stages. First, images with MI above a fold-specific threshold were assigned to radiologist review. Second, among images that passed the MI gate, low calibrated malignancy probabilities received a No-FNA suggestion and high probabilities received an FNA recommendation. Images with intermediate probabilities were also assigned to radiologist review. Consequently, image-based recommendations were issued only for low-MI cases in conservative rule-out or rule-in regions.

Predictive variance and MSP-derived uncertainty were evaluated as secondary uncertainty-score comparators using risk-coverage analysis. MI remained the primary deferral score because the central study question was whether disagreement among independently trained models could identify cases for which automated image-based triage suggestions should be withheld.

After internal cross-validation, a separate TN5000 deployment model was trained and frozen for TN3K evaluation. Because TN3K differed from TN5000 in source, class distribution, annotation format, and ROI extraction, this evaluation was interpreted as a domain-shift analysis of threshold transportability rather than evidence of deployment-ready external generalization.

**Table 2**
Summary of datasets used in this study.

| Dataset | Role | Annotation type | Benign (n, %) | Malignant (n, %) | Total images (n) |
|---|---|---|---|---|---|
| TN5000 | Development and internal evaluation | Bounding boxes and labels | 1426 (28.52%) | 3574 (71.48%) | 5000 |
| TN3K | External evaluation | Segmentation masks and labels | 378 (61.56%) | 236 (38.44%) | 614 |

### 3.3. Nested backbone selection and final model architecture

Backbone selection was repeated independently within each outer fold rather than performed once using the complete TN5000 cohort. Each 4000-image outer-development set underwent three-fold inner architecture selection comparing ConvNeXt-Tiny, EfficientNetV2-S, DenseNet121, ResNet18, and MobileNetV4 under the same preprocessing, augmentation, optimization, early-stopping, and evaluation protocol. Mean inner-fold AUC-ROC determined the primary ranking, and mean inner-fold average precision served as the secondary criterion. The corresponding 1000-image outer test partition was not used for architecture ranking, model fitting, checkpoint selection, or calibration.

ConvNeXt-Tiny was ranked first in all five outer folds (selection frequency, 5/5). The complete nested selection-and-training pipeline yielded a pooled outer-fold AUC-ROC of 0.9405 (95% stratified bootstrap CI, 0.9337-0.9476) and average precision of 0.9724 (95% CI, 0.9680-0.9764), with one outer-fold prediction per TN5000 image. These estimates incorporate image-level data-dependent backbone selection because every outer-test prediction was generated after architecture selection conducted exclusively within the corresponding outer-development partition. TN3K was not used in this procedure. Candidate rankings, outer-fold assignments, image-hash audits, and fold-specific predictions were retained for reproducibility.

A squeeze-and-excitation (SE) channel-attention module [32] was incorporated into the selected classification architecture to recalibrate the pooled channel representation before classification.

Let $X \in \mathbb{R}^{H\times W\times C}$ denote the input feature map. A channel descriptor was obtained by global average pooling:

$$z = GAP\,(X) \tag{1}$$

Channel-wise attention weights were then computed as:

$$S = \sigma\,(W_2\;\;\delta\,(W_1\,z)) \tag{2}$$

where $W_1$ and $W_2$ are learnable weight matrices, δ denotes the ReLU activation function, and σ denotes the sigmoid function. The recalibrated feature map was defined as:

$$\tilde{z} = s \odot z \tag{3}$$

where ⊙ denotes element-wise multiplication. The recalibrated representation was then passed to the classification head. No additional spatial-attention was introduced.

### 3.4. Cross-validation, preprocessing, and ensemble training

The ensemble evaluation used the same five protected outer-fold assignments as the nested architecture-selection procedure. Within each fold, 1000 images (20%) were designated as the untouched outer-test subset before any inner architecture comparison. The remaining 4000 images were divided into four mutually disjoint, role-specific model-development subsets:

a) Training set (n = 3000), used exclusively to optimize model parameters.
b) Early-stopping and checkpoint-selection set (val_es; n = 400), used exclusively for early stopping and selection of the best checkpoint for each ensemble member.
c) Calibration set (val_calib; n = 200), reserved exclusively for fitting member-specific vector-scaling parameters.

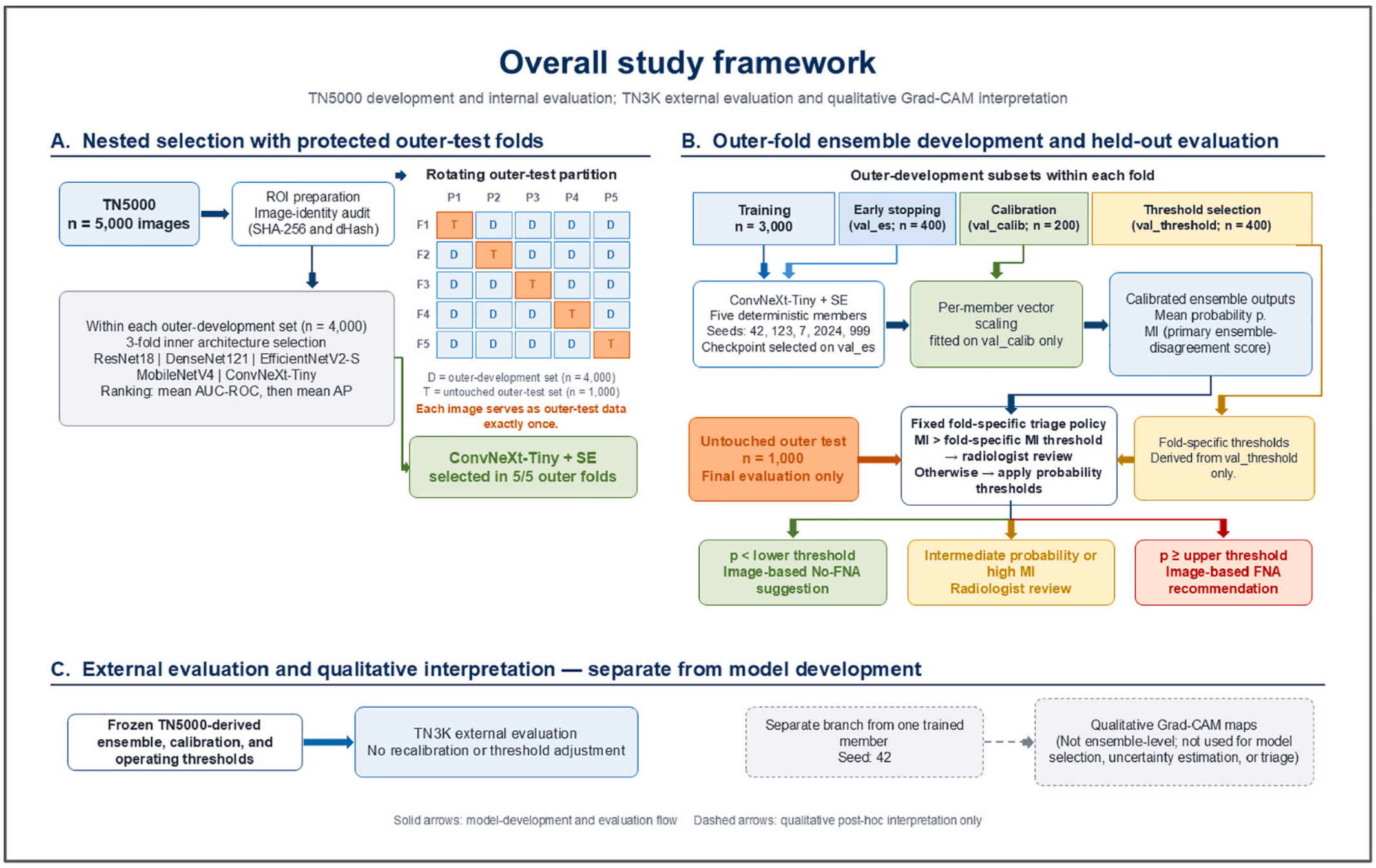

**Fig. 1.** Overall study framework showing nested architecture selection, protected outer-fold ensemble evaluation, selective triage-policy development on TN5000, and external evaluation and qualitative interpretation on TN3K.

d) Threshold-development set (val_threshold; n = 400), reserved exclusively for selecting the fold-specific uncertainty and probability thresholds.

Within each outer fold, the training, early-stopping, calibration, threshold-selection, and outer-test subsets were mutually disjoint. Outer-test images were excluded from architecture ranking, model fitting, checkpoint selection, calibration, and threshold selection and were evaluated only after the fold-specific model and operating policy had been fixed.

ROI images were converted to grayscale, autocontrast-adjusted, resized to 224 × 224 pixels, replicated across three channels, and normalized using ImageNet channel statistics. Training images received stochastic augmentation, including random horizontal flipping, rotation within ±20°, affine transformation, brightness and contrast jittering, Gaussian blurring, and random sharpness adjustment. The val_es, val_calib, val_threshold, and outer-test subsets received deterministic preprocessing only. Class imbalance was addressed during training using an inverse-frequency weighted random sampler with replacement and label-smoothed cross-entropy with a smoothing factor of 0.05. All non-training subsets retained their observed class distributions.

Within each fold, five deterministic ensemble members were trained using the same fold-specific training subset and optimization protocol but different random seeds (42, 123, 7, 2024, and 999). All members were initialized with ImageNet-pretrained ConvNeXt-Tiny weights and optimized using AdamW. The backbone was frozen for the first 15 epochs and unfrozen before epoch 16 for fine-tuning. Training used a mini-batch size of 32 with gradient accumulation over two steps, yielding an effective batch size of 64. The initial learning rate was $5 \times 10^{-4}$, and the fine-tuning backbone learning rate was $1 \times 10^{-5}$. Gradient clipping, learning-rate scheduling, and early stopping were used to stabilize optimization. For each member, the checkpoint with the highest validation average precision on val_es was retained.

Dropout layers were used only for training regularization in the classification head and were disabled during inference. Each image therefore produced one deterministic prediction from each of the five members. These outputs were calibrated and used to calculate the ensemble-mean malignancy probability and MI. Table 3 summarizes the training and deterministic-ensemble configuration.

**Table 3**
Nested cross-validation, training, calibration, and deterministic ensemble configuration.

| Parameter | Configuration |
|---|---|
| Outer-fold design | Five outer folds; 1000 untouched test images per fold |
| Architecture selection | Three-fold inner comparison within each 4000-image outer-development set |
| Fold-specific training subset | 3000 images |
| Early-stopping subset (val_es) | 400 images; checkpoint selection only |
| Calibration subset (val_calib) | 200 images; member-wise vector scaling only |
| Threshold subset (val_threshold) | 400 images; uncertainty and probability thresholds only |
| Selected backbone | ImageNet-pretrained ConvNeXt-Tiny with an added external squeeze-and-excitation block |
| Input image size | 224 × 224 pixels |
| Ensemble size | Five deterministic members |
| Random seeds | 42, 123, 7, 2024, 999 |
| Mini-batch/effective batch size | 32/64 (two-step gradient accumulation) |
| Optimizer | AdamW |
| Initial/fine-tuning learning rate | $5 \times 10^{-4}/1 \times 10^{-5}$ |
| Maximum epochs | 50 |

### 3.5. Probability calibration and ensemble prediction aggregation

Because the strict triage policy depends on probability-based rule-out and rule-in thresholds, vector-scaling calibration was applied before uncertainty estimation, threshold selection, and held-out evaluation. Within each cross-validation fold, each ensemble member was calibrated independently using vector scaling fitted only on the fold-specific val_calib subset.

Let $z_m \in R^2$ denote the uncalibrated two-class logit vector produced by ensemble member (m). The member-specific vector-scaling transformation was:

$$\widetilde{z}_m = w_m \odot z_m + b_m \quad (4)$$

where $w_m$ and $b_m$ are learned two-dimensional scaling and bias vectors, respectively, and ($\odot$) denotes element-wise multiplication. The calibrated malignancy probability for member (m) was then obtained as:

$$p_m = \left[ softmax\left(\widetilde{z}_m\right)\right]_{malignant} \quad (5)$$

For each ensemble member, vector-scaling parameters were fitted exclusively on val_calib by minimizing cross-entropy loss. The fitted parameters were then frozen; no other subset was used to estimate or modify them. Calibration on each outer-test partition was evaluated using expected calibration error with 10 equal-width probability bins and the Brier score.

The calibrated ensemble malignancy probability was calculated as the arithmetic mean of the five calibrated member probabilities:

$$\overline{p} = \left(\frac{1}{M}\right)\sum_{m=1}^{M} p_m \quad (6)$$

where (M = 5). The ensemble-mean probability ($\overline{p}$) was used for rule-out and rule-in decisions, whereas variation among calibrated member probabilities was used for uncertainty estimation.

### 3.6. Mutual information as the primary deferral score

The primary uncertainty score was ensemble predictive mutual information (MI). MI was selected because the central study question was whether disagreement among independently trained deep-learning models could identify thyroid nodule images for which non-review image-based triage should be withheld. MI was calculated from calibrated member probabilities; therefore, it reflects disagreement after fold-specific probability calibration.

For an image, let (p_m) denote the calibrated malignancy probability from ensemble member (m), where ($m = 1, \ldots, M$) and (M = 5). The ensemble-mean probability was ($\overline{p}$), as defined in Section 3.5. For binary classification, predictive entropy was calculated using natural logarithms:

$$H(q) = -q\, log(q) - (1-q)\, log(1-q) \quad (7)$$

Mutual information was calculated as:

$$MI = H(\overline{p}) - \frac{1}{M}\sum_{m=1}^{M} H(p_m) \quad (8)$$

The first term represents uncertainty in the ensemble-mean prediction, whereas the second represents the mean uncertainty of individual ensemble members. MI increases when members produce divergent predictions for the same image. It was therefore interpreted as an operational ensemble-disagreement score and a proxy for epistemic uncertainty, rather than as a complete empirical separation of epistemic and aleatoric uncertainty.

Images with high MI were considered unsuitable for non-review image-based triage and were assigned to radiologist review. Predictive variance and MSP-derived uncertainty were evaluated as secondary uncertainty-score comparators.

### 3.7. Selective prediction evaluation

Selective prediction evaluated whether each uncertainty measure assigned higher scores to images more likely to be misclassified. A selective classifier comprises a base classifier, an uncertainty or confidence score, and a deferral rule [20]. In the retrospective policy evaluation, deferred images were assigned to a separate radiologist-review pathway and were not counted as diagnosed or correctly classified.

Within each held-out outer-test fold, images were ranked from lowest to highest uncertainty. At coverage c, the accepted set comprised the lowest-uncertainty images retained for classification. Risk–coverage and generalized-risk–coverage curves and their corresponding AURC and AUGRC values were calculated independently within each fold [21]. Lower values indicated more effective error ranking. Cross-fold summaries were calculated only after the fold-specific analyses; raw uncertainty values from different fold-specific ensembles and calibrators were not combined into a single ranking.

$$R(c) = \frac{1}{|A_c|}\sum_{i\in A_c} \mathbf{1}(\widehat{y}_i \neq y_i) \quad (9)$$

Generalized risk was defined as the proportion of all images that were both accepted and misclassified:

$$G(c) = \frac{1}{N}\sum_{i\in A_c} \mathbf{1}(\widehat{y}_i \neq y_i) = cR(c) \quad (10)$$

Area under the risk-coverage curve (AURC) and area under the generalized-risk coverage curve (AUGRC) summarize selective and generalized risk across coverage levels:

$$AURC = \int_0^1 R(c)\, dc, AUGRC = \int_0^1 G(c)\, dc \quad (11)$$

Within each outer fold (N = 1000), risk–coverage curves were evaluated at every retained-set size from k = 1 to k = N. Coverage was defined as c = k/N, with images retained in ascending order of uncertainty. AURC and AUGRC were calculated from the resulting full-resolution fold-specific selective-risk and generalized-risk curves.

Secondary analyses compared MI, predictive variance, and MSP-derived uncertainty. Lower AURC and AUGRC values indicated more effective ranking of classification errors. For the rank-normalized pooled sensitivity analysis, each uncertainty measure was converted to a within-fold percentile rank before the five outer folds were concatenated. This transformation preserved within-fold ordering while preventing differences in the raw scales produced by fold-specific ensembles and calibrators from affecting the pooled ranking. Fold-aware stratified bootstrap resampling generated 95% confidence intervals and paired differences between uncertainty measures. Raw uncertainty values were not ranked across folds for inferential comparison.

The complete three-pathway policy was also evaluated separately using MI, predictive variance, and MSP-derived uncertainty. Within each outer fold and for each measure, the uncertainty threshold and lower and upper probability thresholds were derived exclusively from val_threshold using the same nominal-coverage and NPV/PPV target rules. Each fixed policy was then applied once to the corresponding untouched outer-test partition. Fold-aware bootstrap intervals and paired comparisons were calculated for radiologist-review rate, non-review assignment rate, malignant No-FNA rate, No-FNA NPV, and FNA-recommendation PPV. The primary uncertainty measure was not selected on the basis of outer-test performance.

Discrimination and calibration were reported for each held-out outer fold and as mean ± standard deviation across folds. Pooled out-of-fold analyses contained one held-out prediction for each TN5000 image. Operating thresholds were never optimized using pooled outer-test labels.

### 3.8. Strict uncertainty-guided triage policy

The triage policy comprised two sequential decision stages. For fold (f), let ($u_i$) denote the selected uncertainty score for image (i), let ($\tau_u^{(f)}$) denote the corresponding fold-specific uncertainty threshold, and let ($\tau_low^{(f)}$) and ($\tau_high^{(f)}$) denote the lower and upper probability thresholds. MI defined the original primary policy; predictive variance and MSP-derived uncertainty were substituted in secondary comparator policies while the threshold-development procedure remained unchanged.

First, images with ($u_i > \tau_u^{(f)}$) were assigned to radiologist review. Among images that passed the uncertainty gate, the calibrated ensemble probability ($\bar{p}_i$) was used for rule-out and rule-in assignment. Thus, images with intermediate calibrated probabilities were intentionally deferred rather than forced into a No-FNA suggestion or FNA-recommendation pathway. Fig. 1 summarizes the original MI policy, and the same decision structure was used for the predictive-variance and MSP comparator analyses. The complete decision rules are provided in Table 4.

### 3.9. Fold-specific threshold selection and stability analysis

All triage thresholds were derived independently within each cross-validation fold using val_threshold only. The training, val_es, val_calib, and outer-test subsets were not used to select or modify any triage threshold.

The fold-specific MI threshold was selected at nominal MI-gate coverage targets of 50%, 70%, 80%, 90%, and 95%. For each target, the threshold was selected so that the corresponding proportion of val_threshold images with the lowest MI values passed the uncertainty gate. The target therefore represents passage through the MI gate before intermediate-probability deferral; the final non-review assignment rate may be lower because images with intermediate probabilities are assigned to radiologist review.

For each nominal MI-gate coverage target, the lower and upper probability thresholds were derived from the low-MI val_threshold subset. Rule-out threshold selection was restricted to images with an ensemble-mean malignancy probability below 0.50. The lower threshold was the largest candidate for which the strict rule $p < \tau_low$ achieved a No-FNA NPV of at least 98% on val_threshold. If no candidate met the NPV target, no No-FNA assignments were permitted at that operating point.

Rule-in threshold selection was restricted to images with an ensemble-mean malignancy probability of at least 0.50. The upper threshold was the smallest candidate for which the rule $p \geq \tau_high$ achieved an FNA-recommendation PPV of at least 95% on val_threshold. If no candidate met the PPV target, no FNA-recommendation assignments were permitted at that operating point.

The 98% NPV and 95% PPV values were conservative development targets for this retrospective image-based analysis rather than universal clinical standards or guarantees of performance in held-out data. Malignant images were reported according to three mutually exclusive dispositions: No-FNA suggestion, radiologist review, or FNA recommendation. Assignment to radiologist review was not counted as confirmed detection, correct classification, or an observed diagnosis.

Threshold stability was evaluated using 1000 stratified bootstrap resamples of each fold-specific val_threshold subset. Sensitivity analyses evaluated nominal MI-gate coverage targets of 50%, 70%, 80%, 90%, and 95% and jointly varied the NPV/PPV development targets across 0.95/0.90, 0.97/0.925, 0.98/0.95, and 0.99/0.975. Two bootstrap procedures were used. Resampling val_threshold alone quantified threshold stability, whereas joint resampling of val_threshold and the corresponding outer-test partition propagated uncertainty from threshold development and held-out policy evaluation. The primary 80% fold-specific threshold estimates and bootstrap confidence intervals are reported in Table 9; sensitivity analyses at alternative nominal MI-gate coverage targets are reported in Supplementary Section S5 and Supplementary Table S2–S3.

Prevalence sensitivity was evaluated by reweighting the observed fold-specific class-conditional policy outcomes to assumed malignancy prevalences of 5%, 10%, 20%, 30%, 40%, 50%, 60%, and 70%. Prevalence-adjusted No-FNA NPV and FNA-recommendation PPV were calculated while holding the observed class-conditional assignment rates fixed. Stratified outer-test bootstrap resampling generated conditional 95% confidence intervals. This analysis isolates the mathematical effect of prevalence and does not account for spectrum effects, acquisition differences, or changes in class-conditional model performance.

**Table 4**
MI-guided retrospective triage policy.

| Decision stage | Condition | Triage assignment |
|---|---|---|
| Uncertainty gate | ($u_i > \tau_{mi}^{(f)}$) | Radiologist review |
| Rule-out gate | ($u_i \leq \tau_{mi}^{(f)}$) and ($\bar{p}_i < \tau_low^{(f)}$) | No-FNA suggestion |
| Rule-in gate | ($u_i \leq \tau_{mi}^{(f)}$) and ($\bar{p}_i \geq \tau_high^{(f)}$) | FNA recommendation |
| Intermediate-probability region | ($u_i \leq \tau_{mi}^{(f)}$) and ($\tau_low^{(f)} \leq \bar{p}_i < \tau_high^{(f)}$) | Radiologist review |

#### 3.9.1. Ablation analysis

Three locked ablations used the same five protected TN5000 outer folds. The model-aggregation analysis compared the calibrated five-member ConvNeXt-Tiny with SE ensemble with the mean of its five constituent single models; all members were retained, and none was selected using outer-test performance. The architectural analysis compared five-member ensembles with and without the external SE block. The calibration analysis compared vector-scaled and uncalibrated outputs from the with-SE ensemble without additional training. All relevant configurations used identical folds, seeds, preprocessing, optimization, early stopping, calibration, and threshold-development subsets. Vector scaling was fitted only on val_calib, and each configuration's thresholds were estimated independently on val_threshold before one-time evaluation on untouched outer-test data. MSP uncertainty was used for the ensemble and single-model comparison because MI is undefined for a single model; MI was used for the remaining ensemble rows. Paired confidence intervals were obtained by matched within-fold and class resampling for pooled metrics and by resampling fold-level differences for selective outcomes. TN3K was not accessed.

### 3.10. Post-cross-validation deployment training and external domain-shift evaluation

After the five-fold internal evaluation, a separate five-member ConvNeXt-Tiny + SE deployment ensemble was trained on TN5000 using the locked architecture and the same preprocessing, augmentation, optimization, ensemble seeds, deterministic-inference procedure, and vector-scaling calibration framework used for internal development. This deployment ensemble was distinct from the fold-specific cross-validation ensembles and was not used to estimate internal performance.

Before TN3K inference, deployment model weights, member-specific calibration parameters, and TN5000-derived operating policies were fixed. The policy developed at the 50% nominal MI-gate coverage target was designated as the original external policy. For alignment sensitivity analyses, the unchanged TN5000-derived policies corresponding to 70%, 80%, 90%, and 95% coverage targets were also applied to the saved TN3K predictions. TN3K labels were not used for model fitting, checkpoint selection, calibration, uncertainty-score selection, policy derivation, or policy adjustment. The analysis therefore evaluated frozen-policy transportability under dataset shift rather than performance after adaptation to TN3K.

### 3.11. Qualitative Grad-CAM analysis

To enhance interpretability, Grad-CAM [33] was employed to visualize the regions contributing most to the model decision. The Grad-CAM formulation is given by:

$$\alpha_k^c = \frac{1}{Z}\sum_i\sum_j \frac{\partial y^c}{\partial A_{ij}^k} \quad (12)$$

$$L^c_{\text{Grad-CAM}} = \text{ReLU}\left(\sum_k \alpha^c_k A^k\right) \quad (13)$$

where $A^k$ represents the $k$-th feature map, and $\alpha^c_k$ represents the importance weight corresponding to class $c$.

Grad-CAM maps were generated post hoc from the deterministic ensemble member trained with seed 42. The model was placed in evaluation mode, thereby disabling dropout, and Grad-CAM was applied to the final ConvNeXt feature stage using that member's calibrated predicted class as the target. The resulting maps are qualitative single-member visualizations rather than ensemble-level explanations. Grad-CAM did not contribute to architecture selection, model fitting, checkpoint selection, calibration, uncertainty estimation, threshold selection, or triage assignment.

### 3.12. Implementation specifications (Hardware and software environment)

The experimental pipeline was implemented on Kaggle using an NVIDIA Tesla T4 GPU with 16 GB VRAM, a quad-core Intel Xeon CPU, and 30 GB of system RAM. Experiments used Python 3.12.12, PyTorch 2.5, torchvision 0.20, CUDA, scikit-learn 1.4, pandas 2.2, NumPy 1.26, and Pillow in a Linux environment.

To demonstrate how the framework outputs could be presented in an interactive research setting, a web-based prototype displayed ensemble probabilities, member-level predictions, disagreement measures, qualitative Grad-CAM visualizations, and retrospective triage assignments (Supplementary Fig. S4). The prototype was not evaluated in a clinical workflow and is not intended for diagnosis or deployment.

### 3.13. Computational-cost analysis

A computational-cost analysis was performed for the frozen five-member ConvNeXt-Tiny with SE deployment ensemble using all 614 TN3K images. Inference was evaluated on an NVIDIA Tesla T4 GPU using FP32 precision and sequential member execution. Batch sizes of 1 and 32 were assessed in five repeated runs after five warm-up iterations per member. Pipeline wall time included image and mask loading, mask-derived ROI extraction, preprocessing, host-to-device transfer, forward inference, member-specific vector scaling, softmax computation, probability transfer, and ensemble uncertainty aggregation. One-time checkpoint loading was excluded from pipeline latency and measured separately. Latency, throughput, peak GPU memory, and checkpoint storage were recorded. Complete measurement definitions and numerical results are reported in Supplementary Section S4 and Supplementary Table S1.

## 4. Results

### 4.1. Nested backbone architecture selection

ConvNeXt-Tiny ranked first in all five protected outer-development sets. Mean inner-fold AUC-ROC ranged from 0.9266 to 0.9323, and mean inner-fold average precision ranged from 0.9636 to 0.9684. Each selection used only its 4000-image outer-development set; the corresponding 1000-image outer-test partition was excluded from candidate ranking and model selection. Fold-specific outcomes are reported in Table 5.

### 4.2. Fold-specific classification performance

Across the five untouched outer-test folds, the calibrated ensemble achieved a mean AUC-ROC of 0.9397 ± 0.0098 and mean average precision of 0.9717 ± 0.0049. Table 6 reports the fold-specific discrimination, descriptive classification, and calibration estimates. Metrics calculated at a malignancy-probability threshold of 0.50 were descriptive and did not determine the triage operating policies.

### 4.3. Pooled out-of-fold evaluation

#### 4.3.1. Pooled discrimination and calibration

Pooling one held-out prediction per TN5000 image yielded an AUC-ROC of 0.9377 and average precision of 0.9708 (Fig. 2). Fold-specific vector scaling reduced pooled ECE from 0.0288 to 0.0138 and the Brier score from 0.0843 to 0.0827 (Fig. 3). No operating policy was optimized using pooled OOF labels.

#### 4.3.2. Selective-prediction performance

Risk-coverage performance was calculated independently within each untouched outer-test fold. Mean AURC was 0.04503 ± 0.00635 for MI, 0.03616 ± 0.00500 for predictive variance, and 0.03095 ± 0.00504 for MSP uncertainty; the corresponding mean AUGRC values were 0.03044 ± 0.00331, 0.02622 ± 0.00331, and 0.02298 ± 0.00355. Fold-specific estimates and confidence intervals are reported in Table 8, fold-aggregated curves in Fig. 4, and fold-specific curves in Supplementary Fig. S3.

After within-fold percentile normalization, pooled AURC was 0.04499 (95% CI, 0.03946-0.05117) for MI, 0.03611 (0.03183-0.04083) for predictive variance, and 0.03090 (0.02726-0.03492) for MSP uncertainty. Corresponding AUGRC values were 0.03039 (0.02725-0.03366), 0.02617 (0.02344-0.02908), and 0.02293 (0.02069-0.02541). Paired fold-aware bootstrap comparisons confirmed higher, and therefore worse, AURC and AUGRC for MI than for either comparator. MI remained the original primary score because it represented ensemble disagreement, not because it produced the best empirical error ranking.

#### 4.3.3. Direct triage-policy comparison across uncertainty measures

At a common 80% nominal uncertainty-score coverage target, mean radiologist-review rates were 31.14% (95% CI, 26.58%-35.78%) for MI, 29.96% (26.12%-34.18%) for predictive variance, and 32.28% (25.46%-38.48%) for MSP uncertainty. Corresponding malignant No-FNA rates were 0.39% (0.08%-0.70%), 0.22% (0.03%-0.42%), and 0.14% (0.03%-0.28%).

Mean No-FNA NPV was 97.79% for MI, 98.61% for predictive variance, and 98.94% for MSP uncertainty; mean FNA-recommendation PPV was 95.10%, 95.17%, and 95.50%, respectively. Although

**Table 5**
Nested architecture-selection outcomes and protected outer-fold performance.

| Outer fold | Selected backbone | AUC-ROC | AP | Accuracy (%) | Sensitivity (%) | Specificity (%) | MCC | ECE |
|---|---|---|---|---|---|---|---|---|
| 1 | ConvNeXt-Tiny with SE | 0.9277 | 0.9676 | 86.50 | 90.76 | 75.87 | 0.6684 | 0.0371 |
| 2 | ConvNeXt-Tiny with SE | 0.9519 | 0.9775 | 89.70 | 95.66 | 74.74 | 0.7399 | 0.0338 |
| 3 | ConvNeXt-Tiny with SE | 0.9476 | 0.9742 | 90.30 | 96.64 | 74.39 | 0.7551 | 0.0281 |
| 4 | ConvNeXt-Tiny with SE | 0.9328 | 0.9675 | 87.80 | 92.45 | 76.14 | 0.6966 | 0.0351 |
| 5 | ConvNeXt-Tiny with SE | 0.9489 | 0.9780 | 89.30 | 92.87 | 80.35 | 0.7361 | 0.0265 |
| Mean ± SD | Selected in 5/5 folds | 0.9418 ± 0.0108 | 0.9729 ± 0.0052 | 88.72 ± 1.55 | 93.68 ± 2.42 | 76.30 ± 2.38 | 0.7192 ± 0.0357 | 0.0321 ± 0.0046 |

Note: Architecture ranking was conducted independently within each outer-development set. ConvNeXt-Tiny with SE was selected in all five outer folds. Values in the final row are mean ± SD across the protected outer-test partitions.

**Table 6**
Fold-specific calibrated classification and calibration performance across five untouched outer-test folds.

| Metric | Fold 1 | Fold 2 | Fold 3 | Fold 4 | Fold 5 | Mean ± SD |
|---|---|---|---|---|---|---|
| AUC-ROC | 0.9287 | 0.9445 | 0.9478 | 0.9295 | 0.9481 | 0.9397 ± 0.0098 |
| Average precision | 0.9686 | 0.9714 | 0.9753 | 0.9656 | 0.9776 | 0.9717 ± 0.0049 |
| Accuracy (%) | 87.00 | 89.90 | 90.10 | 88.20 | 88.90 | 88.82 ± 1.28 |
| Sensitivity (%) | 93.42 | 93.99 | 96.78 | 95.52 | 94.83 | 94.91 ± 1.32 |
| Specificity (%) | 70.98 | 79.65 | 73.33 | 69.82 | 74.04 | 73.56 ± 3.81 |
| MCC | 0.6720 | 0.7487 | 0.7498 | 0.6999 | 0.7198 | 0.7181 ± 0.0332 |
| ECE | 0.0440 | 0.0258 | 0.0330 | 0.0270 | 0.0202 | 0.0300 ± 0.0091 |
| Brier score | 0.0959 | 0.0748 | 0.0763 | 0.0860 | 0.0803 | 0.0827 ± 0.0086 |

Note: Accuracy, sensitivity, and specificity use a descriptive malignancy-probability threshold of 0.50 and were not used to select the triage thresholds.

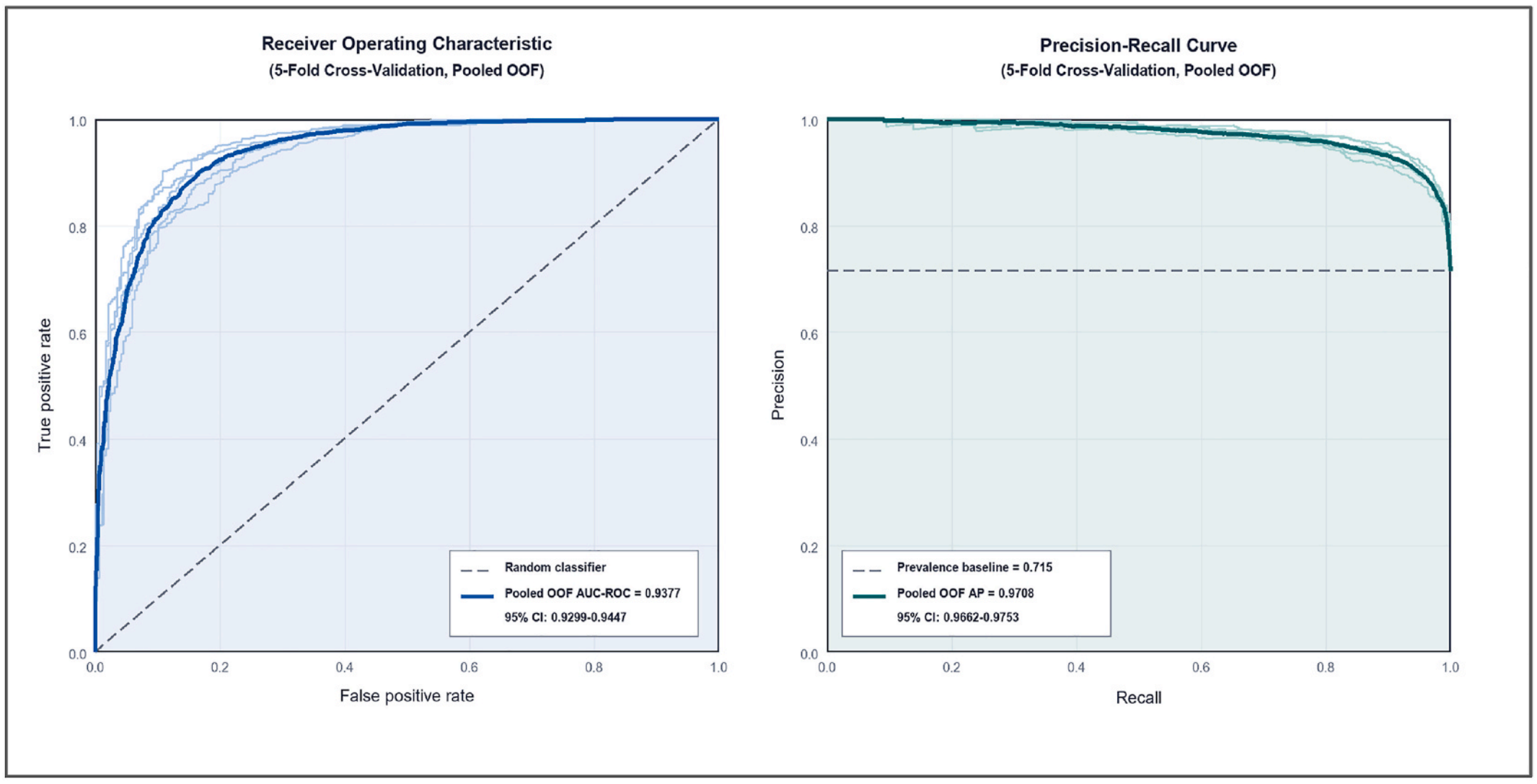

**Fig. 2.** Pooled out-of-fold ROC (A) and precision–recall (B) curves on TN5000. Light lines represent individual folds; dark lines represent pooled calibrated predictions (N = 5000). Pooled AUC–ROC was 0.9377 and average precision was 0.9708; 95% CIs were estimated using 2000 bootstrap resamples.

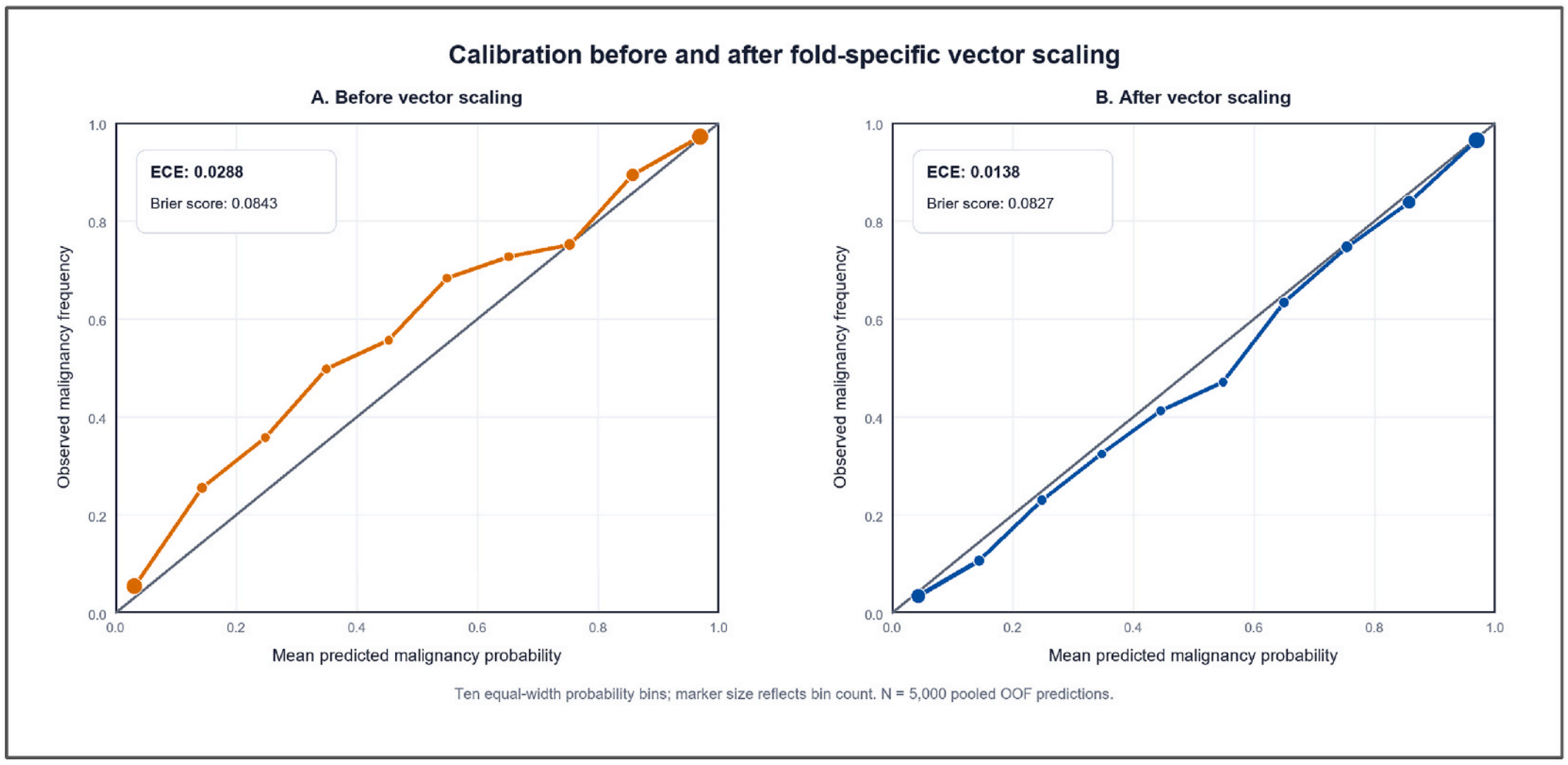

**Fig. 3.** Pooled out-of-fold calibration on TN5000 before (A) and after (B) fold-specific vector scaling fitted exclusively on the corresponding val_calib subset.

predictive variance and MSP uncertainty ranked errors more effectively, no uncertainty measure uniformly optimized radiologist-review burden, non-review assignment rate, and both predictive-value outcomes.

### *4.4. Fold-specific threshold stability and strict triage outcomes*

#### *4.4.1. Fold-specific thresholds and stability analyses*

Table 9 reports the fold-specific threshold estimates and bootstrap confidence intervals at the primary internal-CV operating point, defined by an 80% nominal MI-gate coverage target.

Across folds, MI thresholds were comparatively consistent, whereas

**Table 7**
Primary MI-guided triage outcomes across the five outer-test folds at the 80% nominal MI-gate coverage target.

| Nominal MI-gate coverage target | No-FNA suggestion rate | FNA-recommendation rate | Radiologist-review rate | Malignant dispositions: No-FNA/review/FNA | No-FNA NPV | FNA PPV |
|---|---|---|---|---|---|---|
| 80% | 8.86% ± 5.62% | 60.00% ± 3.62% | 31.14% ± 6.23% | No-FNA: 0.39% ± 0.41% Review: 19.81% ± 4.19% FNA: 79.80% ± 4.26% | 97.79% ± 2.11% | 95.10% ± 0.83% |

Note: Values are mean ± SD across five untouched outer-test folds. The nominal MI-gate coverage target denotes passage through the uncertainty gate before intermediate-probability deferral. Malignant dispositions are reported separately; review assignments are not counted as diagnosed or correctly classified.

**Table 8**
Fold-specific risk-coverage performance across uncertainty measures.

| Uncertainty measure | Outer fold | AURC (95% bootstrap CI) | AUGRC (95% bootstrap CI) |
|---|---|---|---|
| MI | 1 | 0.04481 (0.03487–0.05577) | 0.03345 (0.02669–0.04096) |
| MI | 2 | 0.04768 (0.03421–0.06216) | 0.03001 (0.02311–0.03719) |
| MI | 3 | 0.04446 (0.03151–0.06013) | 0.02816 (0.02183–0.03528) |
| MI | 4 | 0.05282 (0.03907–0.06733) | 0.03410 (0.02634–0.04192) |
| MI | 5 | 0.03539 (0.02640–0.04492) | 0.02646 (0.02046–0.03270) |
| Predictive variance | 1 | 0.03861 (0.03031–0.04773) | 0.02991 (0.02411–0.03612) |
| Predictive variance | 2 | 0.03688 (0.02580–0.04913) | 0.02475 (0.01856–0.03141) |
| Predictive variance | 3 | 0.03412 (0.02419–0.04548) | 0.02417 (0.01851–0.03033) |
| Predictive variance | 4 | 0.04223 (0.03160–0.05378) | 0.02958 (0.02321–0.03632) |
| Predictive variance | 5 | 0.02893 (0.02123–0.03697) | 0.02266 (0.01735–0.02814) |
| MSP uncertainty | 1 | 0.03584 (0.02779–0.04511) | 0.02762 (0.02203–0.03365) |
| MSP uncertainty | 2 | 0.02944 (0.02020–0.04052) | 0.02045 (0.01489–0.02647) |
| MSP uncertainty | 3 | 0.02747 (0.01978–0.03608) | 0.02046 (0.01541–0.02554) |
| MSP uncertainty | 4 | 0.03662 (0.02678–0.04784) | 0.02603 (0.02011–0.03251) |
| MSP uncertainty | 5 | 0.02536 (0.01901–0.03280) | 0.02033 (0.01574–0.02553) |

Note: Metrics were calculated independently within each untouched outer-test fold. Lower values indicate better error ranking. Confidence intervals were obtained using 2000 stratified bootstrap resamples.

**Table 9**
Fold-specific threshold estimates and bootstrap stability at the primary internal-CV operating point (80% nominal MI-gate coverage target).

| Outer fold | MI threshold (95% CI) | Lower probability threshold (95% CI) | Upper probability threshold (95% CI) |
|---|---|---|---|
| 1 | 0.010304 (0.008500–0.012419) | 0.045330 (0.037514–0.293813) | 0.918441 (0.834952–0.974384) |
| 2 | 0.007143 (0.006432–0.008966) | 0.010499 (0.009251–0.266273) | 0.756903 (0.516660–0.885870) |
| 3 | 0.010065 (0.008308–0.011541) | 0.378358 (0.153575–0.486257) | 0.727835 (0.516098–0.909967) |
| 4 | 0.006547 (0.005405–0.008247) | 0.330130 (0.097238–0.488856) | 0.833377 (0.576638–0.894410) |
| 5 | 0.011077 (0.008503–0.012991) | 0.029887 (0.023654–0.412836) | 0.550333 (0.546524–0.920520) |

Note: Point estimates were derived exclusively from the fold-specific val_threshold subset. Confidence intervals were obtained from 1000 stratified bootstrap resamples of that subset.

lower and upper probability thresholds varied substantially and had wide bootstrap intervals. Bootstrap feasibility was at least 99.8%, but outer-test radiologist-review rates ranged from 22.9% to 39.3%. Alternative coverage targets and NPV/PPV target sensitivities are reported in Supplementary Section S5 and Supplementary Tables S2–S3; increasing the NPV/PPV development targets increased review and reduced malignant No-FNA assignment.

### 4.4.2. *Malignant-image dispositions*

At the primary 80% nominal MI-gate coverage target, 443 of 5000 images (8.86%) received a No-FNA suggestion, 3000 (60.00%) received an FNA recommendation, and 1557 (31.14%) were assigned to radiologist review. Mean fold-specific No-FNA NPV was 97.79%, and mean FNA-recommendation PPV was 95.10%; pooled count-based values are distinguished in Table 7 because non-review assignment counts differed across folds.

Among 3574 malignant images, 14 (0.39%) received a No-FNA suggestion, 708 (19.81%) were assigned to review, and 2852 (79.80%) received an FNA recommendation. These mutually exclusive dispositions do not treat review as confirmed detection. They describe retrospective image-based assignments in a malignancy-enriched cohort and are not guarantees, avoided biopsies, or substitutes for guideline-based clinical assessment.

### 4.4.3. *Prevalence sensitivity of predictive values*

At the primary 80% nominal MI-gate coverage target, increasing assumed malignancy prevalence from 5% to 70% changed the fold-mean prevalence-adjusted No-FNA NPV from 99.95% to 97.93% and FNA-recommendation PPV from 29.28% to 94.75%. These conditional sensitivity estimates demonstrate prevalence dependence but do not establish performance under a different clinical case spectrum.

### 4.4.4. *Ablation analysis*

The ensemble and mean single-model results were similar, with modest differences favoring the ensemble. Pooled AUC-ROC was 0.93773 for the ensemble and 0.93498 for the mean single model (difference, 0.00275; 95% CI, 0.00243-0.00310); average precision was 0.97084 and 0.96944, respectively (difference, 0.00140; 95% CI, 0.00118-0.00164). AURC was 0.03095 and 0.03239, respectively (difference, −0.00145; 95% CI, −0.00169 to −0.00117). At the 80% nominal MSP-uncertainty coverage target, No-FNA NPV was 98.94% for the ensemble and 97.85% for the single model; malignant No-FNA rates were 0.14% and 0.26%. Radiologist-review rates were 32.28% and 33.30%, and their difference included zero.

ConvNeXt-Tiny with SE and ConvNeXt-Tiny without SE produced nearly identical results. AUC-ROC was 0.93773 with SE and 0.93775 without SE (difference, −0.00003; 95% CI, −0.00071 to 0.00066); average precision was 0.97084 and 0.97082, respectively (difference, 0.00002; 95% CI, −0.00050 to 0.00053). Confidence intervals for all remaining calibration, risk-coverage, and selective-policy differences included zero, providing no evidence of incremental benefit from the external SE block.

Calibrated and uncalibrated configurations showed calibration-specific trade-offs. ECE was 0.01377 after vector scaling and 0.02883 without calibration (difference, −0.01506; 95% CI, −0.02417 to

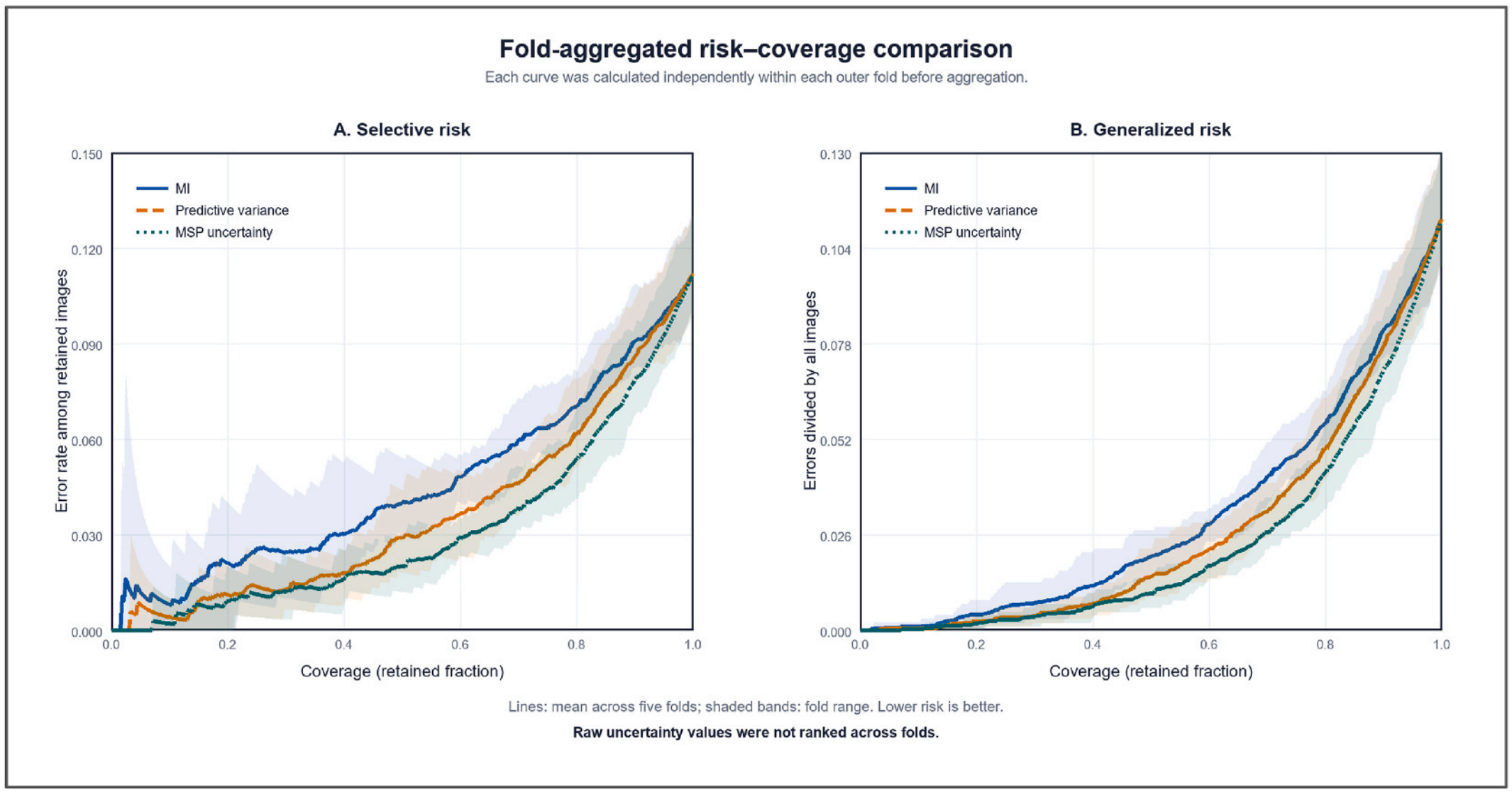

**Fig. 4.** Fold-aggregated risk–coverage analysis for MI, predictive variance, and MSP uncertainty. Lines show five-fold mean selective risk (A) and generalized risk (B); shading indicates fold ranges. Lower values indicate better error ranking.

−0.00368); Brier scores were 0.08267 and 0.08427 (difference, −0.00160; 95% CI, −0.00318 to −0.00006). AUC-ROC was 0.93773 for the calibrated configuration and 0.93928 for the uncalibrated configuration (difference, −0.00156; 95% CI, −0.00279 to −0.00020), whereas the average-precision difference included zero. Under independently fitted 80% MI-guided policies, radiologist-review rates were 31.14% and 33.34%, respectively (difference, −2.20 percentage points; 95% CI, −3.92 to −0.66). Differences in No-FNA NPV, FNA-recommendation PPV, and malignant No-FNA rate included zero. Table 10 summarizes the pooled results for each ablation configuration and the corresponding

**Table 10**
Ablation results on the held-out TN5000 outer-test partitions.

Pooled predictive performance

| Configuration | AUC-ROC | Average precision | Brier score | ECE |
|---|---|---|---|---|
| **Ensemble** | **0.93773** | **0.97084** | **0.08267** | **0.01377** |
| **Single model** | **0.93498** | **0.96944** | **0.08448** | **0.01519** |
| Difference (Ensemble - Single model; 95% CI) | 0.00275 (0.00243 to 0.00310) | 0.00140 (0.00118 to 0.00164) | −0.00181 (−0.00193 to −0.00169) | −0.00142 (−0.00536 to 0.00236) |
| **ConvNeXt-Tiny with SE** | **0.93773** | **0.97084** | **0.08267** | **0.01377** |
| **ConvNeXt-Tiny without SE** | **0.93775** | **0.97082** | **0.08279** | **0.01425** |
| Difference (with SE - without SE; 95% CI) | −0.00003 (−0.00071 to 0.00066) | 0.00002 (−0.00050 to 0.00053) | −0.00012 (−0.00060 to 0.00036) | −0.00048 (−0.00407 to 0.00348) |
| **Calibrated** | **0.93773** | **0.97084** | **0.08267** | **0.01377** |
| **Uncalibrated** | **0.93928** | **0.97150** | **0.08427** | **0.02883** |
| Difference (Calibrated - Uncalibrated; 95% CI) | −0.00156 (−0.00279 to −0.00020) | −0.00066 (−0.00182 to 0.00054) | −0.00160 (−0.00318 to −0.00006) | −0.01506 (−0.02417 to −0.00368) |

Selective prediction at the 80% nominal uncertainty-score coverage target

| Configuration | Deferral score | AURC | Radiologist-review rate | No-FNA NPV | FNA PPV | Malignant No-FNA rate |
|---|---|---|---|---|---|---|
| **Model aggregation** | | | | | | |
| **Ensemble** | **MSP** | **0.03095** | **32.28%** | **98.94%** | **95.50%** | **0.14%** |
| **Single model** | **MSP** | **0.03239** | **33.30%** | **97.85%** | **95.48%** | **0.26%** |
| Difference (Ensemble - Single model; 95% CI) | - | −0.00145 (−0.00169 to −0.00117) | −1.02 pp (−2.05 to 0.07) | 1.09 pp (0.42 to 1.66) | 0.02 pp (−0.29 to 0.36) | −0.12 pp (−0.22 to −0.03) |
| **External SE block** | | | | | | |
| **ConvNeXt-Tiny with SE** | **MI** | **0.04503** | **31.14%** | **97.79%** | **95.10%** | **0.39%** |
| **ConvNeXt-Tiny without SE** | **MI** | **0.04589** | **31.72%** | **98.02%** | **94.99%** | **0.28%** |
| Difference (with SE - without SE; 95% CI) | - | −0.00086 (−0.00270 to 0.00112) | −0.58 pp (−2.88 to 1.72) | −0.23 pp (−1.49 to 1.13) | 0.11 pp (−0.73 to 0.97) | 0.11 pp (−0.17 to 0.42) |
| **Vector-scaling calibration** | | | | | | |
| **Calibrated** | **MI** | **0.04503** | **31.14%** | **97.79%** | **95.10%** | **0.39%** |
| **Uncalibrated** | **MI** | **0.04384** | **33.34%** | **97.83%** | **95.22%** | **0.45%** |
| Difference (Calibrated - Uncalibrated; 95% CI) | - | 0.00119 (−0.00287 to 0.00599) | −2.20 pp (−3.92 to −0.66) | −0.04 pp (−1.04 to 0.92) | −0.13 pp (−0.87 to 0.51) | −0.06 pp (−0.42 to 0.25) |

Note: Predictive metrics are pooled out-of-fold estimates. AURC and selective-policy outcomes are means across the five protected outer folds at the 80% nominal uncertainty-score coverage target. MSP uncertainty was used for the ensemble and single-model comparison because MI is not defined for a single model; MI was used for the with-SE, without-SE, calibrated, and uncalibrated configurations. Difference rows are the first configuration minus the second configuration, with 95% confidence intervals. Pooled metric intervals used 2000 paired, fold-and-class-stratified bootstrap resamples; selective-policy and risk-coverage intervals used 10,000 resamples of the five paired fold differences.

paired differences; fold-specific estimates are reported in Supplementary Section S6 and Supplementary Table S4–S6.

### 4.5. Qualitative Grad-CAM visualization

Fig. 5 presents a correctly classified benign image, a correctly classified malignant image, a misclassified image, and a high-MI image. The maps are qualitative single-member illustrations and do not validate sonographic features, establish causal reasoning, or explain the five-member ensemble. Grad-CAM was not used for model selection, calibration, uncertainty estimation, policy determination, or triage.

### 4.6. External discrimination and calibration under dataset shift

Applied unchanged to TN3K (n = 614), the frozen deployment ensemble achieved an AUC-ROC of 0.7870 (95% CI, 0.7478-0.8237) and average precision of 0.7254 (0.6690-0.7792; Fig. 6). At a descriptive probability threshold of 0.50, accuracy was 66.8%, sensitivity 82.2%, specificity 57.1%, PPV 54.5%, NPV 83.7%, and MCC 0.3878 (Fig. 7). ECE was 0.1899 and the Brier score was 0.2281, demonstrating substantial calibration degradation under dataset shift.

### 4.7. External MI-guided triage and alignment sensitivity under dataset shift

The originally designated external policy, developed at the 50% nominal MI-gate coverage target, was applied unchanged to TN3K. Only 126 of 614 images (20.5%) passed the frozen MI gate. After intermediate-probability deferral, 6 images (1.0%) received a No-FNA suggestion, 94 (15.3%) received an FNA recommendation, and 514 (83.7%) were assigned to radiologist review (Fig. 8).

No malignant image received a No-FNA suggestion; the observed NPV was 100.0% (6/6) but was statistically imprecise. Reviewed cases were reported separately and were not counted as diagnosed or correctly classified. The FNA-recommendation pathway contained 72 malignant and 22 benign images, yielding a PPV of 76.6%, below the TN5000 development target.

To align the external analysis with the primary internal-CV operating point, the unchanged TN5000-derived 80% policy was additionally applied as a sensitivity analysis. It assigned 6 images (1.0%) to No-FNA, 221 (36.0%) to FNA recommendation, and 387 (63.0%) to review; no malignant image received a No-FNA suggestion, whereas FNA-recommendation PPV was 62.4%. Table 11 reports outcomes for all evaluated TN5000-derived coverage targets. These analyses did not use TN3K labels to select or adjust any policy and do not supersede the originally designated 50% external policy.

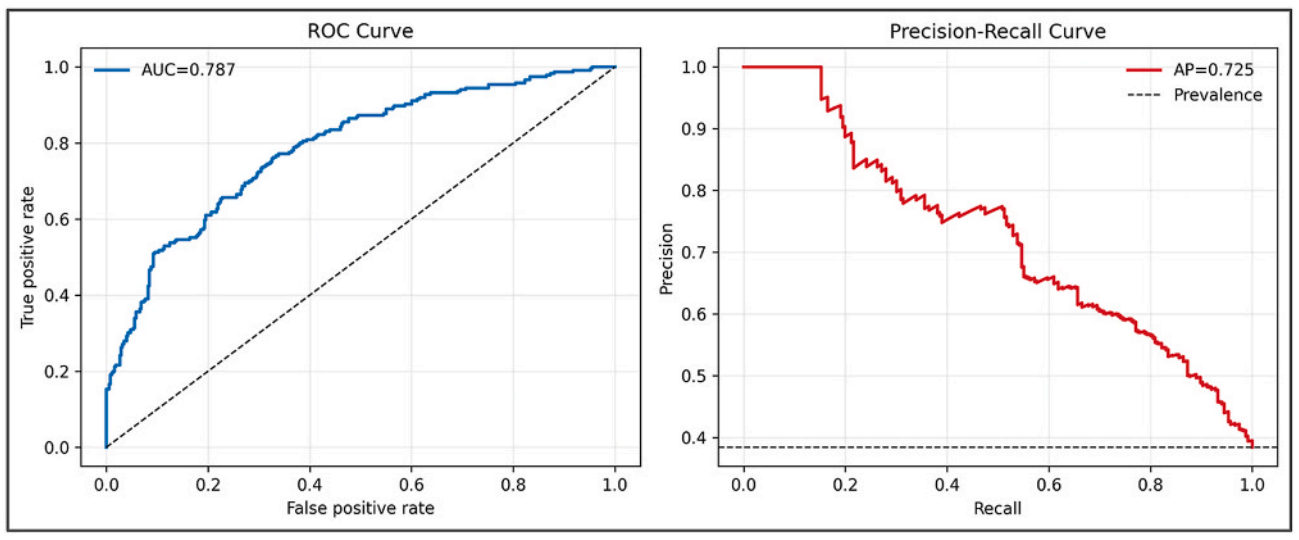

**Fig. 6.** AUC-ROC and Precision-Recall (PR) curves in external TN3K evaluation.

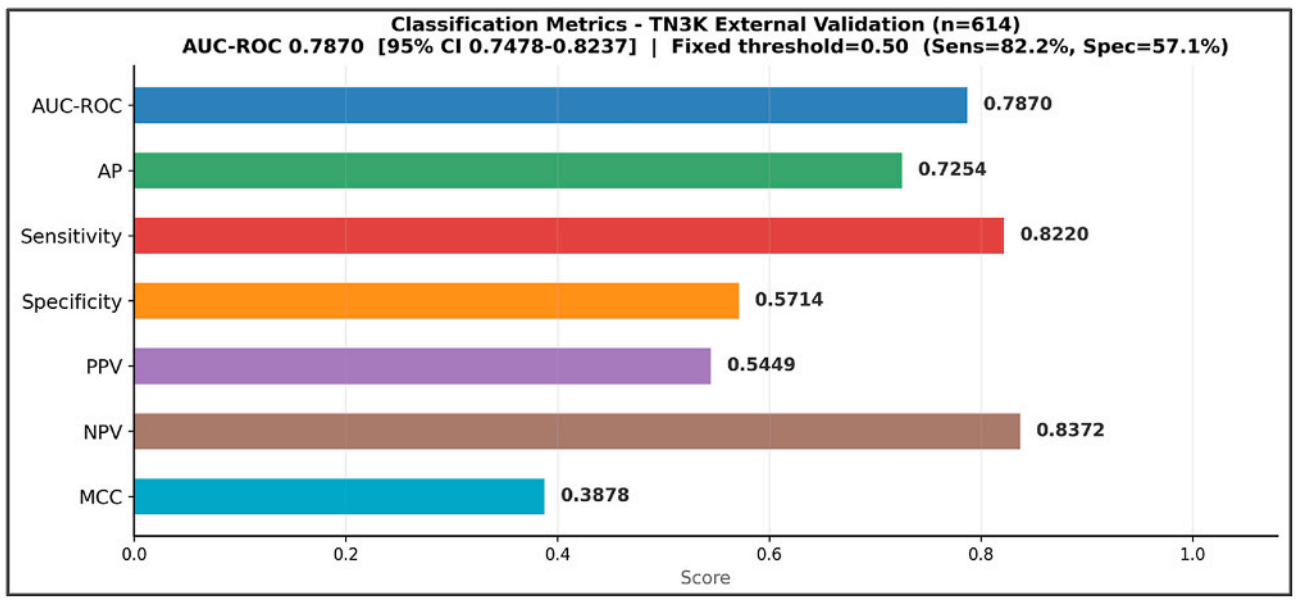

**Fig. 7.** External discrimination and fixed-threshold classification performance on TN3K.

### 4.8. Computational cost

Across five repetitions, batch-size-1 pipeline latency was 44.64 ± 0.61 ms/image and throughput was 22.41 ± 0.31 images/s. At batch size 32, amortized pipeline time was 31.40 ± 6.99 ms/image and throughput was 32.99 ± 6.47 images/s; median amortized time was 28.46 ms/image (range, 25.81-42.33). Peak allocated GPU memory was 130.1 MiB at batch size 1 and 504.2 MiB at batch size 32. The five checkpoints occupied 536.15 MiB, and one-time sequential loading required approximately 2.51-2.60 s. These measurements characterize the tested implementation and hardware rather than prospective clinical workflow performance. Complete computational results are provided in Supplementary Section S4 and Supplementary Table S1.

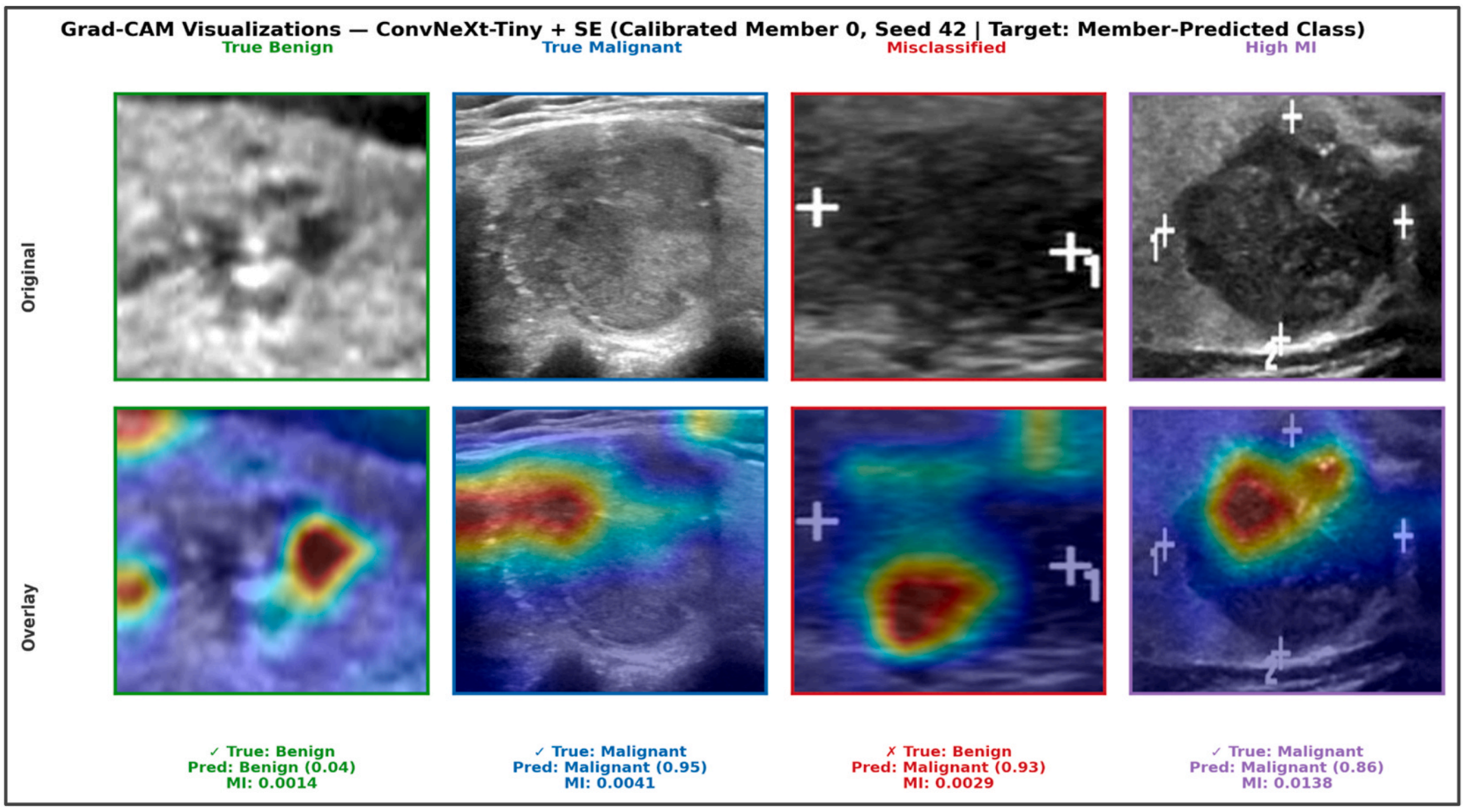

**Fig. 5.** Qualitative Grad-CAM examples showing a true benign, true malignant, misclassified, and high-MI ROI. Rows show the original ROI and Grad-CAM overlay.

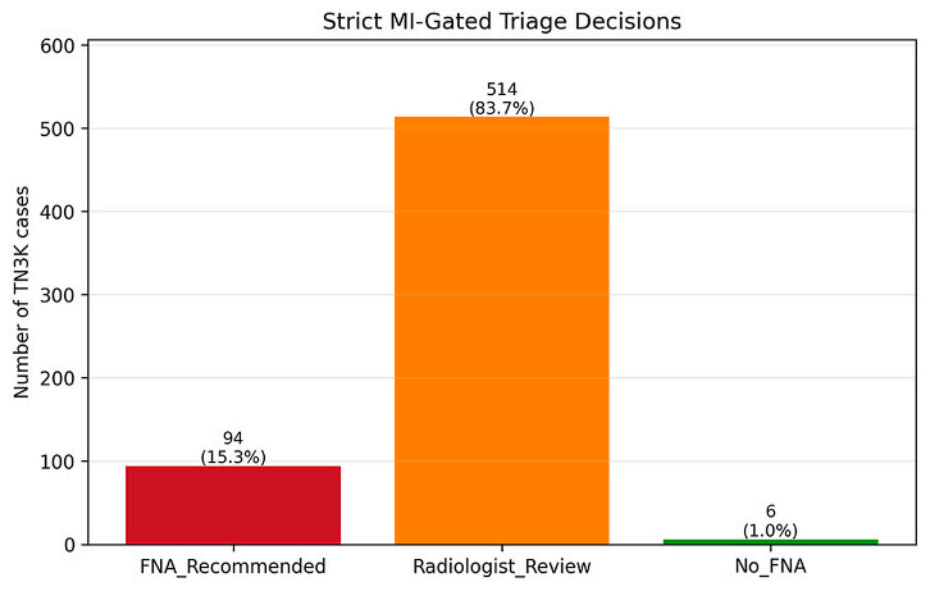

**Fig. 8.** External strict MI-guided triage assignments on TN3K using fixed TN5000-derived thresholds.

## 5. Discussion

This study evaluated a calibrated deterministic deep ensemble as a selective-prediction framework for annotation-assisted thyroid nodule ROI classification. Nested architecture selection, mutually disjoint role-specific development subsets, and protected outer-test partitions separated architecture ranking, model fitting, early stopping, calibration, threshold development, and evaluation. The internal results showed that strong image-level discrimination could coexist with low calibration error; however, the image-level design does not establish patient-level independence or clinical utility.

MI was designated as the primary uncertainty score in the original analysis because the study hypothesis concerned disagreement among independently trained ensemble members. Predictive variance and MSP-derived uncertainty nevertheless ranked classification errors more effectively in both fold-specific and within-fold rank-normalized analyses. Direct policy comparisons further showed that no measure uniformly optimized radiologist-review rate, non-review assignment rate, malignant No-FNA rate, NPV, and PPV. The hypothesis-driven designation of MI as primary should therefore be distinguished from empirical superiority, which was not observed.

The ablation analyses clarified the contributions of individual framework components. The five-member ensemble modestly improved discrimination, calibration, negative log-likelihood, and MSP-based selective-risk ranking relative to the mean performance of the individual members. MSP-derived uncertainty was used for this comparison because MI cannot be calculated for a single model. Performance with and without the added external SE block was similar, providing no evidence of an incremental benefit under the evaluated conditions. Vector scaling reduced calibration error and the radiologist-review rate at the evaluated MI-guided operating point but did not uniformly improve discrimination or other policy outcomes. These findings support ensembling and calibration for their observed functions without attributing benefit to the external SE block or treating calibration as uniformly beneficial across endpoints.

The three-pathway policy combined an MI gate with an intermediate-probability region, thereby withholding image-based suggestions from cases with elevated ensemble disagreement or intermediate malignancy probabilities. Malignant images were reported according to three mutually exclusive dispositions: No-FNA suggestion, radiologist review, or FNA recommendation. Assignment to radiologist review was reported as a separate unresolved disposition and was not counted as confirmed detection, correct classification, or an observed diagnosis. Because the model omitted nodule size, TI-RADS features, clinical history, laboratory findings, and radiologist assessment, these assignments cannot be interpreted as guideline-based biopsy decisions. Their predictive values also varied with assumed malignancy prevalence, demonstrating their dependence on cohort case mix.

External evaluation provided the principal constraint on interpretation. Applying the frozen TN5000-derived deployment ensemble, calibration parameters, and originally designated 50% operating policy to TN3K resulted in poorer discrimination and calibration, extensive assignment to radiologist review, and substantially lower FNA-recommendation PPV. The apparent 100% No-FNA NPV was based on only six assignments and was therefore statistically imprecise. Applying alternative TN5000-derived nominal MI-gate coverage targets without adaptation changed the distribution between review and FNA-recommendation assignments but did not restore the 95% development PPV target. At the 80% alignment target, the lower radiologist-review rate was accompanied by lower FNA-recommendation PPV. These findings demonstrate limited transportability of the TN5000-derived calibration and operating thresholds under external dataset shift.

The computational benchmark documented inference latency, throughput, GPU-memory use, and checkpoint storage for the frozen five-member ensemble under the specified Tesla T4, FP32, and sequential-evaluation configuration. These implementation-specific measurements do not establish computational feasibility in clinical environments, where hardware, image transfer, preprocessing, concurrency, and workflow integration may differ.

Selective prediction should therefore be interpreted as a retrospective routing framework rather than evidence of clinical safety. Deferral can prevent forced image-based assignments, but its value depends on whether uncertainty reliably ranks errors, whether the operating policy

**Table 11**
External TN3K triage outcomes across unchanged TN5000-derived nominal MI-gate coverage targets.

| TN5000 nominal MI-gate coverage target | No-FNA suggestion, n (%) | FNA recommendation, n (%) | Radiologist review, n (%) | Malignant dispositions: No-FNA/review/FNA, n | No-FNA NPV | FNA PPV |
|---|---|---|---|---|---|---|
| 50% | 6 (1.0%) | 94 (15.3%) | 514 (83.7%) | 0/164/72 | 100.0% (6/6) | 76.6% (72/94) |
| 70% | 6 (1.0%) | 178 (29.0%) | 430 (70.0%) | 0/120/116 | 100.0% (6/6) | 65.2% (116/178) |
| 80% | 6 (1.0%) | 221 (36.0%) | 387 (63.0%) | 0/98/138 | 100.0% (6/6) | 62.4% (138/221) |
| 90% | 6 (1.0%) | 152 (24.8%) | 456 (74.3%) | 0/119/117 | 100.0% (6/6) | 77.0% (117/152) |
| 95% | 6 (1.0%) | 154 (25.1%) | 454 (73.9%) | 0/117/119 | 100.0% (6/6) | 77.3% (119/154) |

Note: The 50% target defined the original external policy; 80% aligned with the primary internal operating point, and 70%, 90%, and 95% were sensitivity settings. Each fixed TN5000-derived policy was applied to TN3K without recalibration or threshold adjustment. Review assignments were reported separately and were not counted as diagnosed or correctly classified.

remains valid in the target population, and whether deferred cases can be reviewed within the clinical workflow. Multicentre patient-level validation, site-specific recalibration and threshold validation, integration of guideline-relevant variables, and prospective evaluation of diagnostic safety, workload, downstream procedures, patient outcomes, and subgroup performance are required before clinical implementation can be considered.

## 6. Limitations

This study has several limitations. First, it was retrospective and evaluated annotation-assisted ROI classification. TN5000 ROI patches were derived from provided bounding boxes and TN3K ROI patches were derived from provided pixel-wise masks. The framework therefore did not evaluate end-to-end nodule detection, localization, ROI generation, and triage from a complete uncropped ultrasound examination. The observed classification and triage performance may not be preserved when localization errors are introduced.

Second, internal partitioning was performed at the image level because the released TN5000 materials did not provide patient identifiers or an image-to-patient mapping. Although the TN5000 data descriptor reports that one representative image was retained when multiple images represented the same patient and imaging perspective, this could not be independently verified from the released files. SHA-256 screening confirmed 5000 unique content hashes among 5000 images, and perceptual difference-hash screening identified no candidate image pairs at a Hamming-distance threshold of $\leq 2$. Partition-level assertions prevented sample identifiers, exact content hashes, or defined leakage groups from crossing partition boundaries. These controls prevented detected image-level duplication across partitions but could not establish patient-level independence. If distinct images from the same patient were present under different identifiers, residual correlation across partitions may have inflated internal discrimination and calibration estimates.

Third, architecture selection was limited to five ImageNet-pretrained candidate backbones and was conducted entirely within TN5000. The nested outer–inner design prevented each outer-test partition from influencing its fold-specific architecture ranking and therefore addressed selection-related optimism at the image level. However, the results do not establish that ConvNeXt-Tiny is optimal beyond the evaluated candidate set or across institutions, acquisition protocols, and patient populations.

Fourth, TN5000 was a malignancy-enriched image cohort. Predictive values, pathway-assignment rates, and threshold behavior depend on prevalence and case mix and should not be extrapolated to population-screening settings or unselected thyroid-nodule cohorts. The prevalence-sensitivity analysis recalculated NPV and PPV across assumed malignancy prevalences while holding the observed class-conditional assignment rates fixed. It therefore did not account for spectrum effects, acquisition differences, referral patterns, or changes in class-conditional model performance. The model also omitted variables that ordinarily influence thyroid-nodule management, including nodule size, TI-RADS features, patient characteristics, laboratory findings, longitudinal imaging, and clinician judgment. The No-FNA and FNA-recommendation pathways were retrospective image-based dispositions and not guideline-based decisions to omit or perform biopsy.

Fifth, the fold-specific probability thresholds varied despite the use of an independent threshold-development subset and bootstrap sensitivity analyses. In some folds, the 400-image val_threshold subset contained few low-probability candidates satisfying the high-NPV rule, resulting in wide bootstrap intervals and substantial variation in the number of No-FNA suggestions. These findings limit claims of threshold stability and reinforce the need for larger development cohorts and site-specific threshold validation.

Sixth, external evaluation was limited to a single TN3K cohort and therefore could not characterize performance across the full range of clinical sites, ultrasound systems, acquisition protocols, operators, and patient populations. Differences in annotation format and ROI preparation contributed to the observed dataset shift, but their individual effects could not be isolated. The small number of external No-FNA suggestions made the corresponding NPV estimate highly imprecise. Alternative frozen TN5000-derived nominal MI-gate coverage targets were evaluated as sensitivity analyses and were not locally validated operating policies.

Seventh, the study did not evaluate prospective clinical utility. Subsequent radiologist assessments and final diagnoses were unavailable for cases assigned to radiologist review. The study also did not measure changes in radiologist workload, biopsy utilization, time to diagnosis, patient outcomes, cost-effectiveness, or clinician acceptance.

Eighth, computational performance was measured only on a Tesla T4 GPU using FP32 precision, sequential ensemble-member evaluation, and warm-cache conditions. Batch-size-32 wall-time measurements were variable. The reported latency, throughput, and memory requirements may not generalize to other hardware, inference engines, numerical precisions, storage conditions, or integrated clinical systems. Computational benchmarking therefore does not establish prospective workflow feasibility.

## 7. Conclusion and future work

This study evaluated a calibrated five-member ConvNeXt-Tiny ensemble for selective prediction in annotation-assisted thyroid nodule ROI classification. Nested architecture selection and protected image-level cross-validation showed strong internal discrimination and lower calibration error after vector scaling. MI served as a hypothesis-driven ensemble-disagreement score, although predictive variance and MSP-derived uncertainty ranked classification errors more effectively. The three-pathway policy reported No-FNA suggestion, radiologist review, and FNA recommendation as mutually exclusive dispositions, without counting review assignments as confirmed detection. The associated predictive values remained dependent on prevalence and cohort case mix.

On TN3K, calibration deteriorated substantially, and the frozen TN5000-derived operating policy showed limited transportability. The originally designated 50% policy resulted in extensive assignment to radiologist review, while alternative TN5000-derived nominal MI-gate coverage targets did not remove the need for target-site validation. Selective prediction should therefore be considered a research framework for withholding unreliable image-based suggestions rather than a guideline-based biopsy decision tool. Multicentre patient-level validation, site-specific recalibration and threshold validation, integration of guideline-relevant variables, and prospective workflow evaluation are required before clinical implementation can be considered.

## CRediT authorship contribution statement

**Md. Sadibul Hasan Sadib:** Writing – original draft, Validation, Methodology, Formal analysis, Conceptualization. **Md. Mohayminul Mukit:** Writing – review & editing, Writing – original draft, Visualization, Validation, Methodology, Formal analysis. **Rahmatul Kabir Rasel Sarker:** Writing – review & editing, Validation, Supervision, Investigation. **Tahmid Alam Tamim:** Writing – review & editing, Validation, Formal analysis. **Md. Nazmul Hasan:** Writing – review & editing, Validation, Investigation. **Md. Monir Hossain Shimul:** Writing – review & editing, Visualization, Validation, Supervision, Investigation.

## Consent to participate

Not applicable.

## Clinical trial number

Not applicable.

## Ethics approval

Not applicable. The study used publicly available, de-identified secondary data.

## Consent for publication

Not applicable.

## Availability of data and materials

The datasets used in this study (TN5000 and TN3K) are publicly available and have been described in the Methodology section. The code and experimental results are available at: Code & Experimental Results (https://drive.google.com/drive/folders/1u-5cuPmBJrdZoCxguxjOWcrqLKe9GmCH).

## Declaration of generative AI in scientific writing

Generative AI tools were used only for language editing and manuscript refinement. All scientific content, analyses, interpretations, and final responsibility remain with the authors.

## Funding statement

The authors received no specific funding for this work.

## Declaration of competing interest

The authors declare that they have no known competing financial interests or personal relationships that could have appeared to influence the work reported in this paper.

## Acknowledgements

Not applicable.

## Appendix A. Supplementary data

Supplementary data to this article can be found online at https://doi.org/10.1016/j.ibmed.2026.100517.